\documentclass[12pt]{article}

\usepackage{amsfonts,amssymb,amsmath}
\usepackage{graphicx}
\DeclareGraphicsExtensions{.pdf,.png,.jpg}
\begin{document}
\title{\bf Well-protected quantum state transfer in a dissipative spin chain}
\author{ Naghi Behzadi $^{a}$ \thanks{E-mail: n.behzadi@tabrizu.ac.ir} ,
Abbas Ektesabi $^{b}$ and
Bahram Ahansaz $^{b}$
\\ $^a${\small Research Institute for Fundamental Sciences, University of Tabriz, Iran,}
\\ $^b${\small Physics Department, Azarbaijan Shahid Madani University, Iran.}} \maketitle

\begin{abstract}
\noindent
{In this work, a mechanism for improving the quantum state transfer efficiency in a spin chain, which is in contact with a dissipative structured reservoir, is investigated. The efficiency of the method is based on the addition of similar non-interacting auxiliary chains into the reservoir. In this regard, we obtain the exact solution for the master equation of the spin chain in the presence of dissipation. It is found out that entering more auxiliary chains into the reservoir causes, in general, the better improvement of the fidelity of state transfer along the mentioned chain. Furthermore, it is reveal that the protocol has better efficiency for a chain with longer length. Therefore, by this method, quantum state transfer along a linear chain with an arbitrary number of qubits, can be well-protected against the dissipative noises.}
\end{abstract}

\section{Introduction}

The high-fidelity transmission of quantum states from one location to another in a quantum network through a quantum channel is an important task in quantum information processing. This is so because any performance of a quantum information processing task inside a quantum computer needs to exchange quantum information between distant nodes. Among the various physical systems, quantum spin chains are the best-known ones that can serve as quantum channels. After the pioneer work of Bose $\cite{Bose}$, in which an unmodulated ferromagnetic spin chain with nearest neighbor Heisenberg interaction was proposed as a channel for short range quantum communication, various theoretical frameworks were proposed to increase the transmission fidelity in quantum state transfer (QST) $\cite{Osborne, Wojcik, Lyakhov}$ and even to achieve perfect state transfer (PST) in spin chains [5-13]. Interesting situations arise if one assume to have individual control on the nearest neighbor couplings in the spin chain, in particular, PST can be achieved by properly engineering and modulating of these couplings $\cite{Christandl, Christandl1, Albanese}$.

On the other hand, since any real system is inevitably subjected to its surrendering environment, achieving QST with high fidelity in the presence of noise and dissipation effects is an outstanding challenge in quantum channels.
So it would be important to consider possible methods to minimize or eliminate this unwanted effects on the QST efficiency, as considered recently in $\cite{Hu, Zwick, Morigi}$.

In this paper, we propose a theoretical approach to achieve to high fidelity transmission of a quantum state in a linear spin chain which is in contact with a dissipative structured reservoir. It is assumed that the PST is achievable for the isolated spin chain due to the same pre-engineered nearest-neighbor couplings discussed in the Refs. \cite{Albanese, Chakrabarti}. The performance of the method is based on the enterance of other similar auxiliary spin chains, without direct interaction with each other, into the reservoir. In this direction, we provide the analytical solution for the dynamics of the chains immersed in the reservoir. It is found out that increasing the number of auxiliary chains leads to access to a high fidelity state transfer. Furthermore, it is figured out that for a chain with more qubits we have a better decoupling of the unitary dynamics of the chain from the dissipation, which means that the protocol has better efficiency for the chains with longer length.

In the following sections, we first study the PST in a spin chain according to the Refs. \cite{Albanese, Chakrabarti}. In the next step, the exact dynamics of the system in the presence of dissipative noises is obtained and consequently, the mechanism for protection of QST process against the dissipative noises, in the spin chain, is investigated. Finally, the paper is ended by a brief conclusion.

\section{Perfect State transfer for an isolated spin chain}
We consider a set of $M$ identical qubits on a linear chain with nearest-neighbor $XY$ coupling. The Hamiltonian of the system is given by
\begin{eqnarray}
\hat{H}=\omega_{0}\sum_{j=0}^{M-1}\sigma_{j}^{+}\sigma_{j}^{-}+\sum_{j=0}^{M-2 }J_{j}(\sigma_{j}^{+}\sigma_{j+1}^{-}+\sigma_{j}^{-}\sigma_{j+1}^{+})
\end{eqnarray}
where $\omega_{0}$ is the transition frequency and $J_{j}$ is the coupling strength between the qubits located at site $j$ and $j+1$.
The lowering operator $\sigma_{j}^{-}=(\sigma_{j}^{+})^{\dagger}=|\textbf{0}\rangle\langle j|$ describes decay from the excited state $|j\rangle\equiv|\underbrace{000...1_{j}...0}_{M}\rangle$ into the ground state $|\textbf{0}\rangle\equiv|\underbrace{000...0...0}_{M}\rangle$ where $|j\rangle$ describes the state in which there is an excitation in the qubit located at the site $j$ $(j=0, 1, 2, ..., M-1)$. The states $\{|j\rangle\}$ are considered as a set of basis for the single excitation subspace of the spin chain called as canonical or standard basis.

The matrix representation of the Hamiltonian (1) in this basis takes the following form
\begin{eqnarray}
\hat{H}=\begin{pmatrix}
\omega_{0}&J_{0}&0&...&0&0\\
J_{0}&\omega_{0}&J_{1}&...&0&0\\
0&J_{1}&\omega_{0}&...&0&0\\
.&.&.&.&.&.\\
.&.&.&.&.&.\\
.&.&.&.&.&.\\
0&0&0&...&\omega_{0}&J_{M-1}\\
0&0&0&...&J_{M-1}&\omega_{0}
\end{pmatrix}.
\end{eqnarray}
This Hamiltonian is real and symmetric, so from the spectral theorem $\cite{Golub}$ it can be written as
\begin{eqnarray}
\hat{H}=\hat{U}\hat{D}\hat{U}^{T}
\end{eqnarray}
where $\hat{D}$ is a diagonal matrix and $\hat{U}$ is an orthogonal one as
\begin{eqnarray}
\hat{D}=\mathrm{diag}(E_{0},E_{1},...,E_{M-1}),
\qquad\qquad
\hat{U}\hat{U}^{T}=\hat{U}^{T}\hat{U}=1,
\end{eqnarray}
by noting that $E_{j}$s denote energy eigenvalues of $\hat{H}$, and $T$ as the transpose operation. The columns of the matrix $\hat{U}$ are the eigenvectors of $\hat{H}$ as
\begin{eqnarray}
|\phi_{l}\rangle=\begin{pmatrix}
U_{0l}\\
U_{1l}\\
.\\
.\\
.\\
U_{(M-1)l}
\end{pmatrix}=\sum_{j=0}^{M-1}U_{jl}|j\rangle
,\quad
(l=0,1,...,M-1)
\end{eqnarray}
with $\hat{H}|\phi_{l}\rangle=E_{l}|\phi_{l}\rangle$. From the orthogonality of $\hat{U}$, the inverse relation reads
\begin{eqnarray}
|j\rangle=\sum_{l=0}^{M-1}U_{jl}|\phi_{l}\rangle.
\end{eqnarray}
The dynamics of the system is described by the unitary time evolution operator $\mathcal{\hat{U}}(t)\equiv exp(-i\hat{H}t)$. We assume that there is an excitation at site $0$ ($j=0$) of the chain at $t=0$ which we desire to transfer it to site $M-1$. After a certain time $t$ the system evolves to the state $\mathcal{\hat{U}}(t)|0\rangle$ (or $\mathcal{\hat{U}}(t)|j=0\rangle=\mathcal{\hat{U}}(t)|\underbrace{100...0...0}_{M}\rangle$) which, in general, is a superposition of various standard states $|j\rangle$s. So the transition amplitude for an excitation in transferring from one end of the chain to the other end is given by the transfer fidelity or \emph{fidelity of state transfer} as follows
\begin{eqnarray}
f_{0, M-1}(t)=\langle M-1|\mathcal{\hat{U}}(t)|0\rangle.
\end{eqnarray}
Substituting Eqs. (5) and (6) into Eq. (7), gives
\begin{eqnarray}
f_{0, M-1}(t)=\sum_{l=0}^{M-1}U_{0,l}U_{M-1,l}e^{-iE_{l}t}.
\end{eqnarray}
The situation of PST at time $t$ from one end of the chain to the other one occurs when $|f_{0,M-1}(t)|=1$. This can be accomplished through the proper choice of coupling strength $J_{j}$ between the adjacent qubits. If the coupling strengths of the spin chain are chosen to be uniform, the PST does not occur for chains containing more than three qubits $\cite{Christandl1, Albanese}$. However, it is known that by choosing the coupling strengths as $J_{j}=\sqrt{(j+1)(M-j-1)}$ ($j=0, 1, ..., M-1$), the PST is achievable $\cite{Christandl1, Chakrabarti,Jeugt}$.
It turns out that in this sense the eigenvectors of $\hat{H}$ (the columns of $\hat{U}$) are related to the well-known \emph{Krawtchouk polynomial} $\cite{Albanese, Chakrabarti}$ as
\begin{eqnarray}
|\phi_{l}\rangle=\sum_{j=0}^{M-1}U_{jl}|j\rangle=\sum_{j=0}^{M-1}\tilde{K}_{l}(j)|j\rangle,
\end{eqnarray}
where $\tilde{K}_{l}(j)$ is the orthonormal Krawtchouk function defined as
\begin{eqnarray}
\tilde{K}_{l}(j)\equiv\frac{\sqrt{w(j)}K_{l}(j)}{\sqrt{d_{l}}},
\end{eqnarray}
where $K_{l}(j)$ is the Krawtchouk polynomial $\cite{Koekoek, Nikiforov, Ismail}$ of degree $l$ $(l=0,1,...,M-1)$ in the variable $j$, with parameter $0<p<1$, written as follows
\begin{eqnarray}
K_{l}(j)=F_{1}(-j,-l;-M+1;\frac{1}{p}).
\end{eqnarray}
The function $F_{1}$ is the classical hypergeometric series $\cite{Bailey}$ and in this case it is a terminating series because of the appearance of the negative $-l$ as a numerator parameter. The $w(j)$ is the weight function in $j$, and $d_{l}$ is a function depending on $l$ as
\begin{eqnarray}
w(j)=\begin{pmatrix}
M-1\\
j\\
\end{pmatrix}p^{j}(1-p)^{M-1-j}
,\qquad
d_{l}=\frac{1}{\begin{pmatrix}
M-1\\
l\\
\end{pmatrix}}\big(\frac{1-p}{p}\big)^{l},
\end{eqnarray}
where for the aim of this paper we set $p=\frac{1}{2}$.
Also the corresponding energy eigenvalues of $\hat{H}$ are
\begin{eqnarray}
E_{l}=M-1-2l.
\end{eqnarray}
Let us consider that there is an excitation at first qubit of the chain while the others are in ground state.
By substituting of Eqs. (9) and (13) into Eq. (8), one can compute the fidelity of state transfer for an excitation from site $0$ to $M-1$ as
\begin{eqnarray}
|f_{0, M-1}(t)|=|\mathrm{sin}(t)|^{M-1}.
\end{eqnarray}
Eq. (14) gives the perfect state transfer between two ends of the chain with transfer time $t=\frac{\pi}{2}$.

\section{Protection process in the presence of dissipation}

In this stage we consider the spin chain as an open quantum system in which the efficiency of state transfer process is degraded due to the existence of interaction between the chain and a dissipative structured reservoir. In other words, all of the qubits in the chain is contained in a common reservoir. We introduce the protection process by considering other $N-1$ auxiliary similar chains with $M$ spin such that each of these chains also is involved in the above mentioned reservoir (see figure 1). It is assumed that there is no direct interaction between the chains. The Hamiltonian of the whole system reads as
\begin{eqnarray}
\begin{array}{c}
\hat{\bar{H}}=\omega_{0}\sum_{i=1}^{N} \sum_{j=0}^{M-1}\sigma_{i,j}^{+}\sigma_{i,j}^{-}
\\\\
+\sum_{i=1}^{N}\sum_{j=0}^{M-1}\sqrt{(j+1)(M-j-1)}\big(\sigma_{i,j}^{+}\sigma_{i,j+1}^{-}+\sigma_{i,j}^{-}\sigma_{i,j+1}^{+}\big)
\\\\
+\sum_{k}\omega_{k}b_{k}^{\dag}b_{k}+\sum_{i=1}^{N}\sum_{j=0}^{M-1}\sum_{k}\big(g_{k,j}\sigma_{i,j}^{+}b_{k}+g_{k,j}^{*}\sigma_{i,j}^{-}b_{k}^{\dag}\big),
\end{array}
\end{eqnarray}
where $\omega_{0}$ is the transition frequency, $b_{k}$ ($b_{k}^\dag$) is the annihilation (creation) operator for the $k$th field mode with frequency $\omega_{k}$. In the above equation, we have introduced the site-dependent coupling strength $g_{k,j}$ as the coupling constant between the $k$th field mode and the $j$th qubit located at site $j$ of the chains defined as
\begin{eqnarray}
g_{k,j}=g_{k}\tilde{K}_{0}(j),
\end{eqnarray}
where $\tilde{K}_{0}(j)$ has been defined in Eq. (10) (for $l=0$). Taking the site-dependent coupling strength $g_{k,j}$ as Eq. (16), leads to the exact solution of the master equation for the dynamics of the system. It should be noted that taking each of the $\tilde{K}_{l}(j)$ for $l=0, 1, 2, ..., M-1$, also gives the exact solution of the master equation. For the lowering operator we have  $\sigma_{ij}^{-}=(\sigma_{ij}^{+})^{\dagger}\equiv|\bar{\textbf{0}}\rangle\langle{i,j}|$ where $|\bar{\textbf{0}}\rangle\equiv|\underbrace{000...0...0}_{N\times M}\rangle$ and $|i,j\rangle\equiv|\underbrace{000...1_{_{i,j}}...0}_{N\times M}\rangle=|i\rangle\otimes|j\rangle$. In fact, $|i,j\rangle$ indicates that there exist an excitation in the $j$th site of the $i$th chain with $j=0, 1, ..., M-1$ and $i=1, 2, ..., N$. The states $\{|i,j\rangle\}$ can be considered as a set of basis for the single excitation subspace of $N$ similar chains each of which has $M$ identical qubits. As the previous section, by considering the unitary transformation $\hat{\bar{U}}=I_{N\times N}\otimes \hat{U}$ ($\hat{U}$ is defined in Eq. (9) as a $M\times M$-matrix with elements $U_{j,l}=\tilde{K}_{l}(j)$), we can transform the Hamiltonian (15) to the diagonal form $\hat{\bar{H}}=\hat{\bar{U}}\hat{\bar{D}}\hat{\bar{U}}^{^{T}}$ where
\begin{eqnarray}
\hat{\bar{D}}=\mathrm{diag}(E_{j}^{i})
\qquad\text{with}\qquad
E_{j}^{i}=(M-1)-2j,
\end{eqnarray}
where $j=0,1,...,M-1$ and $i=1,2,...,N$. The columns of the matrix $\hat{\bar{U}}$ are the eigenvectors of $\hat{\bar{H}}$ and related to the Krawtchouk polynomials as follows
\begin{eqnarray}
|\Phi_{l}^{i}\rangle\equiv|i\rangle\otimes|\phi_{l}\rangle=\sum_{j=0}^{M-1}U_{j,l}|i,j\rangle=\sum_{j=0}^{M-1}\tilde{K}_{l}(j)|i,j\rangle.
\end{eqnarray}
From the orthogonality of $\hat{\bar{U}}$, the inverse relation follows as
\begin{eqnarray}
|i,j\rangle=\sum_{l=0}^{M-1} \tilde{K}_{l}(j)|\Phi_{l}^{i}\rangle.
\end{eqnarray}
So the Hamiltonian (15) in the basis $\{|\Phi_{l}^{i}\rangle\}$ takes the following form
\begin{eqnarray}
\hat{H}=\sum_{i=1}^{N}\sum_{l=0}^{M-1}(\omega_{0}+E_{l}^{i})|\Phi_{l}^{i}\rangle \langle\Phi_{l}^{i}|+
\sum_{k}\omega_{k} b_{k}^{\dag}b_{k}+\sum_{i=1}^{N}\sum_{k}\big(g_{k}\Xi^{i+}b_{k}+g_{k}^{*}\Xi^{i-}b_{k}^{\dag}\big),
\end{eqnarray}
where
\begin{eqnarray}
\Xi^{i+}=(\Xi^{i-})^\dagger \equiv\sum_{j=0}^{M-1}\tilde{K}_{0}(j)|i,j\rangle\langle\bar{\mathbf{0}}|=|\Phi_{0}^{i}\rangle\langle\bar{\mathbf{0}}|.
\end{eqnarray}
It is clear from the Eq. (21) that the interaction of the system with the reservoir takes place collectively only through the eigenstate $|\Phi_{0}^{i}\rangle$ and therefore the $N(M-1)$ eigenstates of the system are decoupled from the reservoir. On the other hand if we choose each of the $\tilde{K}_{l}(j)$ for $l=1, 2, 3, ..., M-1$ in Eq. (16), then the coupling of the system to the reservoir is provided only through the related $|\Phi_{l}^{i}\rangle$ and therefore the other $N(M-1)$ eigenstates are decoupled from the reservoir.

Now we consider the dynamics of the system by noting to the point that at initial time $t=0$, there exist only a single excitation in one of the chains and the other $N-1$ chains along with the reservoir are in their respective ground states. Let us assume that the initial state can be written, in general, as follows
\begin{eqnarray}
|\psi(0)\rangle=C(0)|\bar{\mathbf{0}}\rangle_{S}|0\rangle_{E}+\sum_{i=1}^{N}\sum_{l=0}^{M-1}C_{l}^{i}(0)|\Phi_{l}^{i}\rangle_{S}|0\rangle_{E}.
\end{eqnarray}
Since the Hamiltonian conserves the number of excitations in the system, the time-evolved state $|\psi(t)\rangle$ is
\begin{eqnarray}
|\psi(t)\rangle=C(0)|\bar{\mathbf{0}}\rangle_{S} |0\rangle_{E}+\sum_{i=1}^{N}\sum_{l=0}^{M-1}C_{l}^{i}(t)|\Phi_{l}^{i}\rangle_{S}|0\rangle_{E}+\sum_{k} C_{k}(t)|\bar{\mathbf{0}}\rangle_{S}|1_{k}\rangle_{E},
\end{eqnarray}
where $|1_{k}\rangle_{E}$ denotes the state of the reservoir with only one excitation in the $k$th mode. The time-dependent coefficients $C_{l}^{i}(t)$ and $C_{k}(t)$ are determined from the schr\"{o}dinger equation $i\frac{d}{dt}|\psi(t)\rangle=\hat{\bar{H}}|\psi(t)\rangle$, as follows
\begin{eqnarray}
\frac{d C_{0}^{i}(t)}{d t}=-i(\omega_{0}+E_{0}^{i})C_{0}^{i}(t)-i\sum_{k}g_{k}C_{k}(t)
,\qquad
\frac{d C_{l\neq0}^{i}(t)}{d t}=-i(\omega_{0}+E_{l}^{i})C_{l\neq0}^{i},
\end{eqnarray}

\begin{eqnarray}
\frac{d C_{k}(t)}{d t}=-i\omega_{k}C_{k}(t)-i\sum_{j=1}^{N}g_{k}^{*}C_{0}^{i}(t).
\end{eqnarray}
A convenient way to solve the above equations is to use the following redefinitions
\begin{eqnarray}
\begin{array}{c}
\tilde{C}_{l}^{i}(t)=e^{i(\omega_{0}+E_{l}^{i})t}C_{l}^{i}(t),\quad l=0, 1, 2, ..., M-1,\\\\
\tilde{C}_{k}(t)=-e^{i\omega_{k}t}C_{k}(t).
\end{array}
\end{eqnarray}
Now by substituting Eq. (26) into Eqs. (24) and (25), we can obtain the following differential equations
\begin{eqnarray}
\frac{d \tilde{C}_{0}^{i}(t)}{d t}=-i\sum_{k}g_{k}e^{i(\omega_{0}+E_{0}^{i}-\omega_{k})t}\tilde{C}_{k}(t)
,\qquad
\frac{d \tilde{C}_{l\neq0}^{i}(t)}{d t}=0,
\end{eqnarray}

\begin{eqnarray}
\frac{d \tilde{C}_{k}(t)}{d t}=-ig_{k}^{*}e^{-i(\omega_{0}+E_{0}-\omega_{k})t}\sum_{i=1}^{N}\tilde{C}_{0}^{i}(t).
\end{eqnarray}
Integrating Eq. (28) and substituting it into Eq. (27) gives the integro-differential equation
\begin{eqnarray}
\frac{d \tilde{C}_{0}^{i}(t)}{d t}=-\int_{0}^{t} f(t-t')\sum _{i=1}^{N}\tilde{C}_{0}^{i}(t')dt',
\end{eqnarray}
where the correlation function$f(t-t') $ is related to the spectral density $J(\omega)$ of the reservoir by
\begin{eqnarray}
f(t-t')=\int_{0}^{\infty}d\omega J(\omega)e^{-i(\omega_{0}+E_{0}^{i}-\omega)(t-t')}.
\end{eqnarray}
Here, the structure of the common reservoir can be described by an effective Lorentzian spectral density of the form
\begin{eqnarray}
J(\omega)=\frac{1}{2\pi}\frac{\gamma_{0}\lambda}{(\omega-\omega_{0})^2+\lambda^2},
\end{eqnarray}
where $\lambda$ is the spectral width, $\gamma_{0}$ the coupling strength, and $\omega_{0}$ is the central frequency of the reservoir which is equal the transition frequency of qubits. Using the Laplace transformation and its inverse, we can obtain a formal solution for $\tilde{C}_{0}^{i}(s)$ as
\begin{eqnarray}
\begin{array}{c}
\tilde{C}_{0}^{i}(t)=e^{-(\lambda-iE_{0}^{i})t/2}\Big(\mathrm{cosh}(\frac{Dt}{2})+\frac{\lambda-iE_{0}^{i}}{D}\mathrm{sinh}(\frac{Dt}{2})\Big)\tilde{C}_{0}^{i}(0)\\\\
+\bigg(1-e^{-(\lambda-iE_{0}^{i})t/2}\Big(\mathrm{cosh}(\frac{Dt}{2})+\frac{\lambda-iE_{0}^{i}}{D}\mathrm{sinh}(\frac{Dt}{2})\Big)\bigg)\frac{\sum_{l\neq 0} \tilde{C}_{0}^{i}(0)-\tilde{C}_{l}^{i}(0)}{N},
\end{array}
\end{eqnarray}
where
\begin{eqnarray}
D=\sqrt{(\lambda-iE_{0}^{i})^{2}-2\gamma_{0}\lambda N}.
\end{eqnarray}
Also we can obtain $\tilde{C}_{l\neq0}^{i}(t)=\tilde{C}_{l\neq0}^{i}(0)$ by considering the second part of Eq. (27). Then by using Eq. (26), we can acquire the formal solution for the probability amplitudes $C_{l}^{i}(t)$s ($l=0, 1, 2, ..., M-1$).

Now we return to the basis $\{|i,j\rangle\}$ and obtain the $|\psi(t)\rangle$ in (23), in terms of this basis as follows
\begin{eqnarray}
|\psi(t)\rangle=\xi(0)|\bar{\mathbf{0}}\rangle_{S}|0\rangle_{E}+\sum_{i=1}^{N}\sum_{j=0}^{M-1}\xi_{j}^{i}(t)|i,j\rangle_{S}|0\rangle_{E}
+\sum_{k} C_{k}(t)|\bar{\mathbf{0}}\rangle_{S}|1_{k}\rangle_{E},
\end{eqnarray}
where
\begin{eqnarray}
\xi_{j}^{i}(t)=\sum_{l=0}^{M-1}\sqrt{\frac{w(l)}{d_{l}}}K_{l}(j)C_{l}^{i}(t),
\end{eqnarray}
is the probability amplitude for the excitation of the $j$th qubit located in the $i$th chain.

In this step, we impose the initial condition in such way that only the qubit at site $0$ of the $1$th chain is initially excited, i.e. $ \xi_{0}^{i=1}(0)\neq0$ and $ \xi_{j\neq0}^{i\neq1}(0)=0$ with $|\xi(0)|^{2}+|\xi_{0}^{i=1}(0)|^{2}=1$. In fact by the protection process introduced in this paper, we expect that the the quantum state $|\psi\rangle=\xi(0)|0\rangle_{S}+\xi_{0}^{i=1}(0)|1\rangle$ prepared initially at one end of a given chain, for example the $1$th one, can be enabled to transfer to the other end of this chain with a high fidelity of state transfer. This is equivalent to the evolution of the state $|\psi(t=0)\rangle=\xi(0)|\bar{\mathbf{0}}\rangle_{S}+\xi_{0}^{i=1}(0)|\underbrace{1_{_{1, 0}}00...0...0}_{N\times M}\rangle_{S}$ to the target state $|\psi_{tar}\rangle=\xi(0)|\bar{\mathbf{0}}\rangle_{S}+\xi_{0}^{i=1}(0)|\underbrace{000...1_{_{1, M-1}}...0}_{N\times M}\rangle_{S}$, at a certain time $t$ with a considerable fidelity. On the other hand, to obtain the $\xi_{j}^{i=1}(t)$ in Eq. (35), as an explicit function of time corresponding to the given initial condition, we impose this condition on the $C_{l}^{i=1}(t)$s at $t=0$. So due to the unitarity of $\hat{\bar{U}}$, as denoted in Eq. (18), and using Eq. (35), it is obtained the following relation for ${C}_{l}^{i=1}(0)$ as
\begin{eqnarray}
C_{l}^{i=1}(0)=\sqrt{\frac{w(0)}{d_{l}}}K_{l}(0)\xi_{0}^{i=1}(0).
\end{eqnarray}
Therefore, the probability amplitude for finding the initial excitation, at time $t$, in the qubit located at site $j$ of the $1$th chain is given by
\begin{eqnarray}
\xi_{j}^{i=1}(t)=\chi_{j}^{i=1}(t)\xi_{0}^{i=1}(0),
\end{eqnarray}
where
\begin{eqnarray}
\begin{array}{c}
\chi_{j}^{i=1}(t)=\frac{w(0)}{\sqrt{d_{0}d_{j}}}e^{-i(\omega_{0}+E_{0}^{1})t}\\\\
\times\bigg(\frac{N-1}{N}+\frac{e^{-(\lambda-iE_{0}^{1})t/2}}{N}\Big(\mathrm{cosh}(\frac{Dt}{2})+\frac{\lambda-iE_{0}^{1}}{D}\mathrm{sinh}(\frac{Dt}{2})\Big)\bigg)+\sum_{l=1}^{M-1}\sqrt{\frac{w(0)w(l)}{d_{l}d_{j}}}K_{l}(j)e^{-i(\omega_{0}+E_{l}^{1})t}.
\end{array}
\end{eqnarray}
After tracing out from the state (34) with respect to the degrees of freedom of the structured reservoir and all of the qubits except the the qubit located at the end of the $1$th chain, the reduced density matrix becomes as
\begin{eqnarray}
\rho_{M-1}^{i=1}(t)=\begin{pmatrix}
|\xi_{M-1}^{i=1}(t)|^{2}&\xi_{M-1}^{i=1}(t)\xi^{\ast}(0)\\\\
\xi_{M-1}^{\ast i=1}(t)\xi(0)&1-|\xi_{M-1}^{i=1}(t)|^{2}\\
\end{pmatrix}.
\end{eqnarray}
Consequently, the fidelity between the state (39) and the state $|\psi\rangle$, is obtained as
\begin{eqnarray}
F\big(|\psi\rangle\langle\psi|, \rho_{M-1}^{i=1}(t)\big)=\sqrt{\langle\psi|\rho_{M-1}^{i=1}(t)|\psi\rangle}
\\\nonumber&&\hspace{-60mm}=\sqrt{|\xi(0)|^{2}\Big( 1-2|\xi_{M-1}^{i=1}(t)|^{2}+\xi_{M-1}^{i=1}(t)\xi_{0}^{\ast i=1}(0)+\xi_{M-1}^{\ast i=1}(t)\xi_{0}^{ i=1}(0)\Big)+|\xi_{M-1}^{i=1}(t)|^{2}}.
\end{eqnarray}
Since the $|\bar{\mathbf{0}}\rangle_{S}$ component of the the state $|\psi(0)\rangle$ is invariant under the evolution, it suffices to focus to the choice $\xi(0)=0$ and $\xi_{0}^{ i=1}(0)=1$. Therefore, it is concluded that the fidelity of state transfer for an excitation between two ends of the dissipative spin chain in the presence of other $N-1$ similar auxiliary chains contained in the reservoir is written as
\begin{eqnarray}
|f_{0, M-1}(t)|=|\xi_{M-1}^{i=1}(t)|=|\chi_{M-1}^{i=1}(t)|.
\end{eqnarray}
Figs. (2), (3) and (4), demonstrate the performance of the QST protocol introduced in this paper for the chains of length $M=2, 3, 4$. Fig. 2, shows the QST efficiency represented in terms of the fidelity of state transfer for a two-qubit spin chain, i.e. $M=2$. For this case, in the absence of additional chains, i.e. $N=1$, the fidelity of state transfer in Eq. (41), is strongly affected by the dissipation. It is observed a steady value for the fidelity of state transfer. In fact, the interaction of the two-qubit spin chain with the common reservoir is established only through the respective eigenstate $|\Phi_{0}\rangle=1/\sqrt{2}(|10\rangle+|01\rangle)$, and the other eigenstate $|\Phi_{1}\rangle=1/\sqrt{2}(|10\rangle-|01\rangle)$ is decoupled from the reservoir. Since $|\psi(0)\rangle_{S}=|10\rangle_{S}=1/\sqrt{2}(|\Phi_{0}\rangle+|\Phi_{1}\rangle)$, therefore $|f_{0, 1}(t\rightarrow\infty)|=\frac{1}{2}$, as shown in Fig. 2. On the other hand, in the presence of additional chains (for example $N=50$), it is observed a considerable improvement in the efficiency of state transfer (see Fig. 2). Evidently, whatever $N$ becomes larger the state transfer process in the mentioned chain is better protected against the dissipative noise.

Fig. 3, shows the quantum state transfer along a spin chain with three qubits. In the absence of the additional chains, the initial state is $|\psi(0)\rangle_{S}=|100\rangle_{S}=\frac{1}{2}|\Phi_{0}\rangle+\frac{1}{2}|\Phi_{1}\rangle+\frac{1}{\sqrt{2}}|\Phi_{2}\rangle$,
where $|\Phi_{0}\rangle=\frac{1}{2}(|100\rangle+\sqrt{2}|010\rangle+|001\rangle)$, $|\Phi_{1}\rangle=\frac{1}{2}(|100\rangle-\sqrt{2}|010\rangle+|001\rangle)$ and $|\Phi_{2}\rangle=\frac{1}{\sqrt{2}}(|100\rangle-|001\rangle)$ are the eigenstates of the three-qubit chain. Obviously, $|\psi(0)\rangle_{S}$ has a support on the decoupled subspace spanned by $\{|\Phi_{1}\rangle_{S}, |\Phi_{2}\rangle_{S}\}$, so the interaction of the three-qubit spin chain with the reservoir is possible only through the eigenstate $|\Phi_{0}\rangle$. Therefore, by entering the corresponding three-qubit auxiliary chains ($N=45$), the QST for the three-qubit spin chain can be well-protected against the noise. This procedure can be repeated for the four-qubit chain by considering the four-qubit auxiliary chains ($N=40$), as depicted in Fig. 4.

Consequently, by these observations, the QST protocol has a better efficiency for the spin chains with more qubits.

\section{Conclusions}
In summery, we investigated a mechanism for the protection of the intrinsic PST of a pre-engineered linear spin chain in the presence of dissipative noises. By obtaining the exact dynamics, it was shown that the protection process can be well-controlled through the entering non-interacting auxiliary chains into the structured reservoir and therefore, high fidelity state transmission is achievable in the considered spin chain. Furthermore, it was illustrated that the protocol has better efficiency for the chains with more qubits.

\newpage
Fig. 1. A schematic representation of a spin chain in the presence of, for example, four similar auxiliary chains contained in the reservoir.
\begin{figure}
\centering
\includegraphics[width=280 pt]{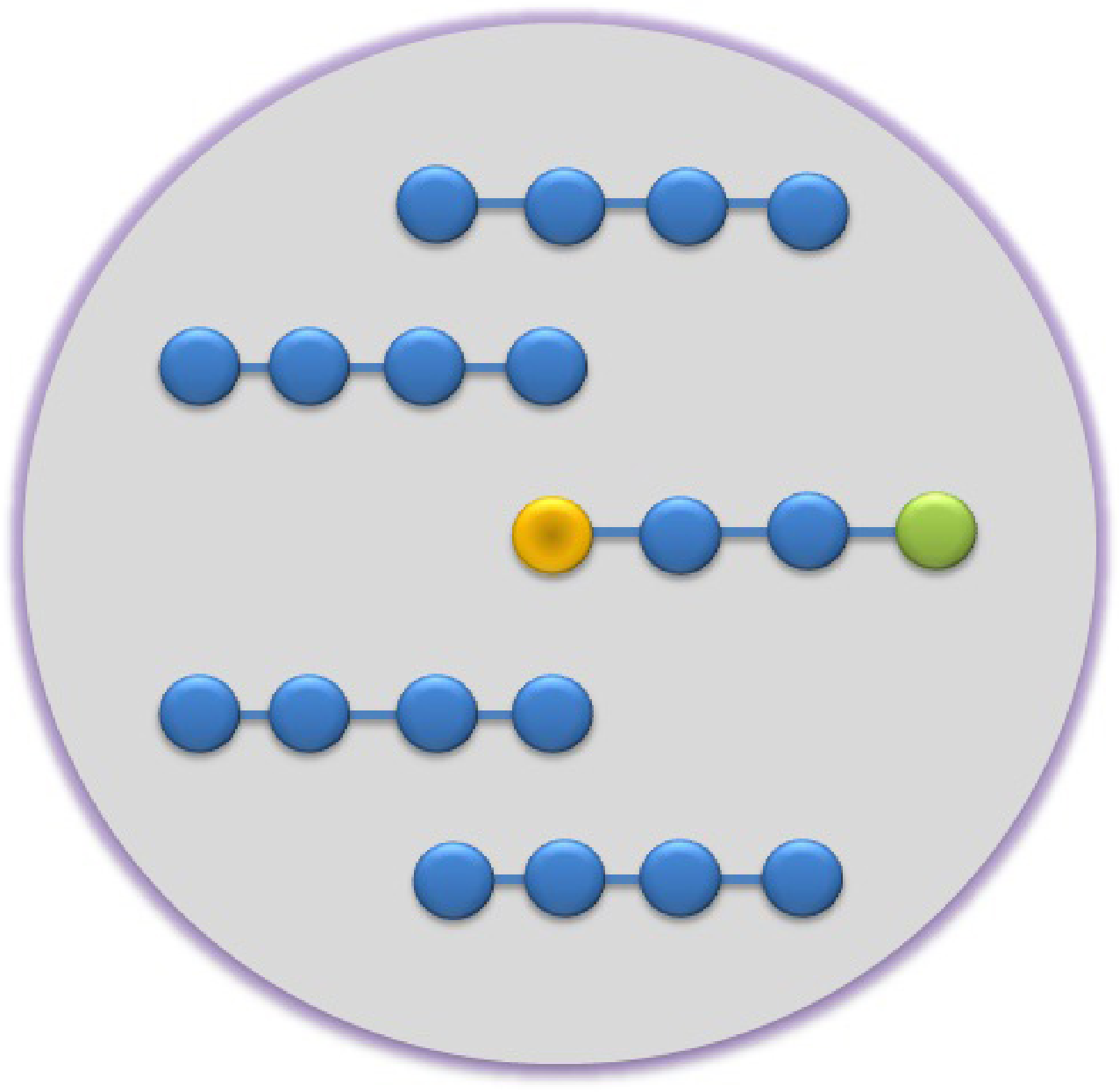}
\caption{} \label{fig1}
\end{figure}

\newpage
Fig. 2. Fidelity of state transfer for the protected ($N=50$) and unprotected ($N=1$) two-qubit spin chain contained in the Lorentzian reservoir. The related parameters are $\lambda=50$ (in units of $\gamma_{0}$) and $\omega_{0}=1$ (in units of $\gamma_{0}$). As illustrated in text, $N=50$ means that there are $49$ similar auxiliary chains in the reservoir for the aim of protection process.
\begin{figure}
\centering
\includegraphics[width=445 pt]{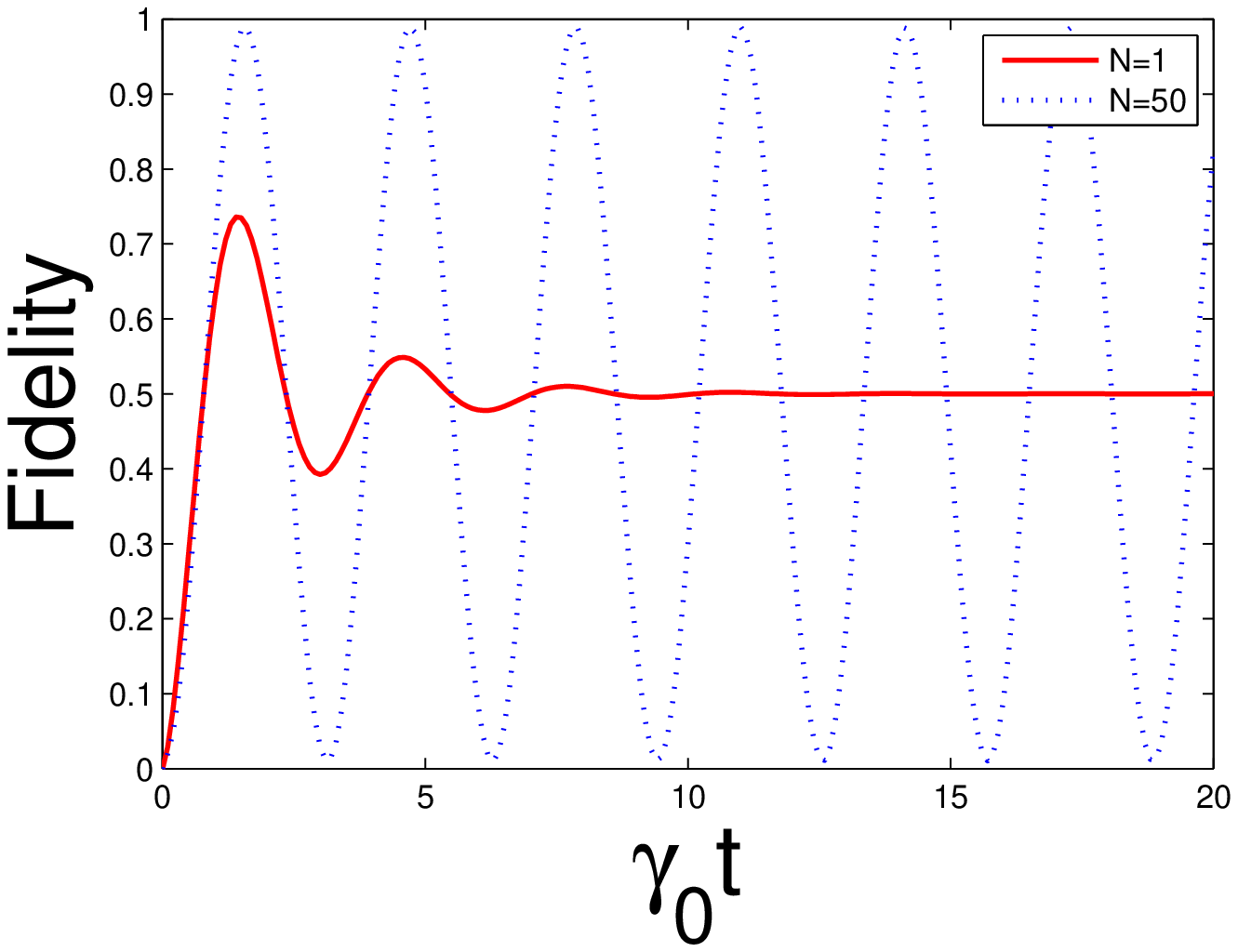}
\caption{} \label{fig2}
\end{figure}

\newpage
Fig. 3. Fidelity of state transfer for the protected ($N=45$) and unprotected ($N=1$) three-qubit spin chain contained in the Lorentzian reservoir. The related parameters are $\lambda=50$ (in units of $\gamma_{0}$) and $\omega_{0}=1$ (in units of $\gamma_{0}$).
\begin{figure}
\centering
\includegraphics[width=445 pt]{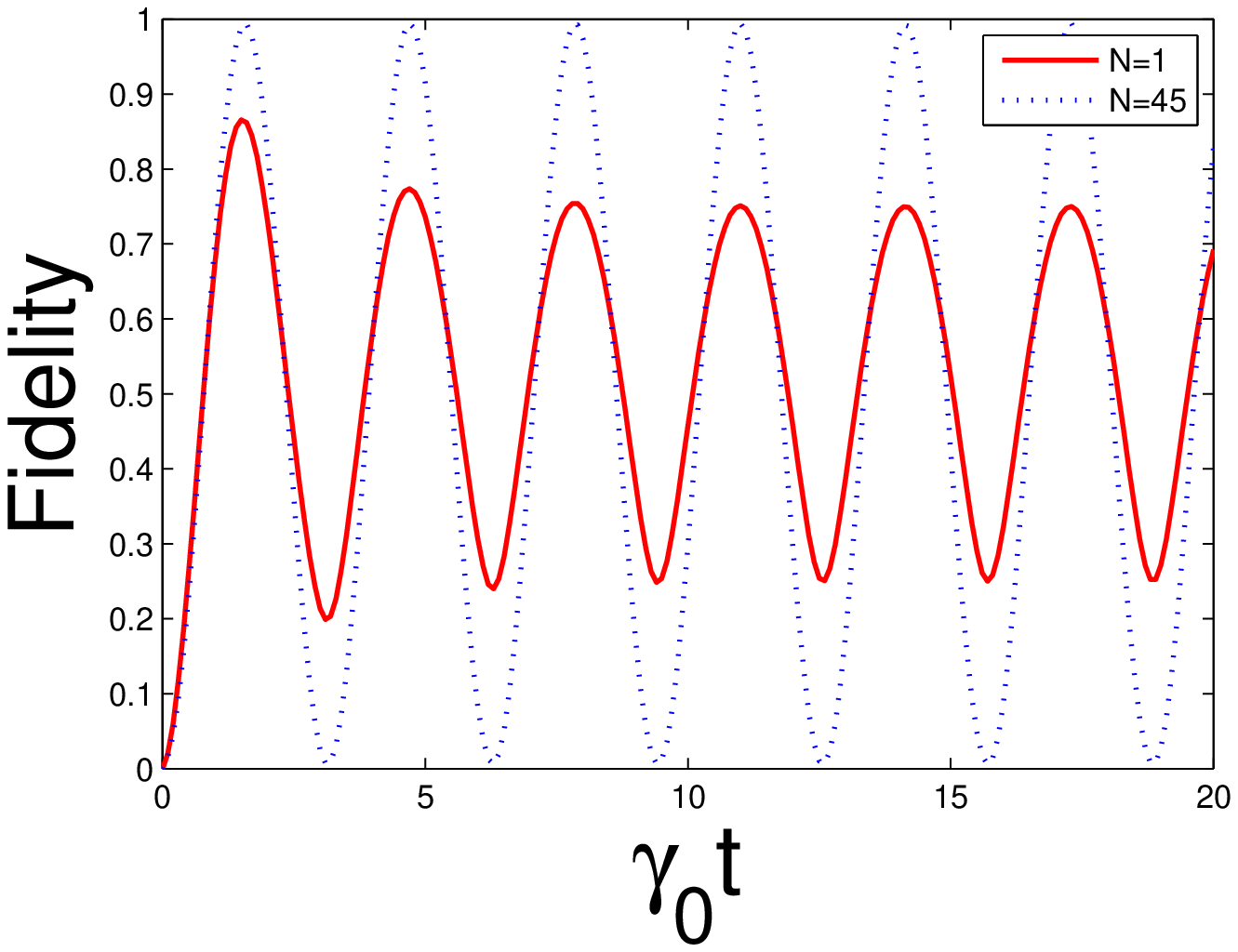}
\caption{} \label{fig3}
\end{figure}

\newpage
Fig. 4. Fidelity of state transfer for the protected ($N=40$) and unprotected ($N=1$) four-qubit spin chain contained in the Lorentzian reservoir. The related parameters are $\lambda=50$ (in units of $\gamma_{0}$) and $\omega_{0}=1$ (in units of $\gamma_{0}$).
\begin{figure}
\centering
\includegraphics[width=445 pt]{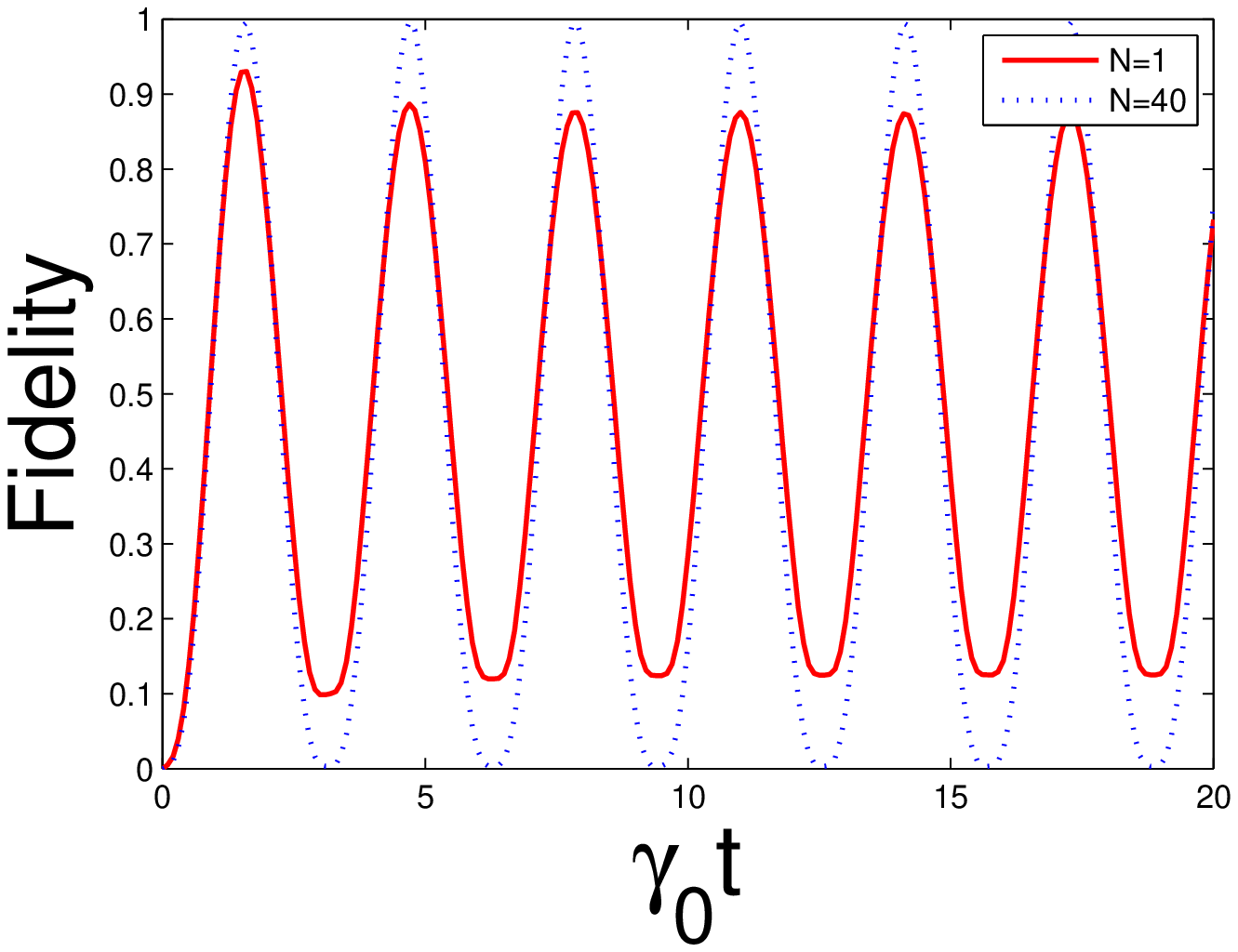}
\caption{} \label{fig4}
\end{figure}
\end{document}